\documentclass[aps,prb,twocolumn,floatfix,footinbib,showpacs,superscriptaddress]{revtex4-1}
\usepackage{graphicx}
\usepackage{amsfonts,amsmath,amssymb}
\usepackage{amsthm}
\usepackage{dsfont,bm}
\usepackage{color}
\usepackage{soul} 
\usepackage{amsbsy}
\usepackage[colorlinks=true,linkcolor=blue,pagecolor=blue,filecolor=blue,menucolor=blue,urlcolor=blue,citecolor=blue,anchorcolor=blue]{hyperref}%
\usepackage{sidecap}
\usepackage{txfonts}
\usepackage{amsbsy}
\usepackage{times} 
\usepackage{dcolumn}

\newcommand{\ua}{\uparrow}
\newcommand{\da}{\downarrow}

\begin{document}

\title{Evolution of Pair Correlation Symmetries and Supercurrent Reversal  in Tilted Weyl Semimetals}
\author{Mohammad Alidoust}
\affiliation{Department of Physics, K.N. Toosi University of Technology, Tehran 15875-4416, Iran}
\author{Klaus Halterman }
\affiliation{Michelson Lab, Physics Division, Naval Air Warfare Center, China Lake, California 93555, USA}

\date{\today} 
\begin{abstract} 
We study the effective symmetry profiles of superconducting pair correlations and 
the flow of charge supercurrent in ballistic Weyl semimetal systems with a
tilted dispersion relation. Utilizing a microscopic method in the ballistic regime and starting from both {\it opposite}-pseudospin (the band degree of freedom) and {\it equal}-pseudospin phonon-mediated opposite-spin electron-electron couplings, we calculate the anomalous 
Green's function to study 
various superconducting pair correlations that Weyl semimetal systems may develop. 
The momentum-space profile reveals that by properly manipulating the parameters of Weyl semimetal systems, 
including the tilting parameter, the effective symmetry class of even-parity $s$-wave (odd-parity $p$-wave) 
superconducting correlations can be converted into a $d$-wave ($f$-wave) symmetry class
that consists of equal-pseudospin and opposite-pseudospin channels. 
We also find that the 
supercurrent in a ballistic Weyl Josephson junction can be made to vanish or switch directions, depending on the 
tilt of the Weyl cones, in addition to the relevant 
parameters characterizing the Weyl semimetal and junction. We show that inversion symmetry 
breaking terms introduce transitions that result in the appearance of self-biased current at zero difference between the macroscopic phases
of the superconducting segments, creating a $\varphi_0$ Josephson state. 
Weyl semimetal systems are shown to 
offer several experimentally tunable parameters to control the 
induction of higher harmonics into the current phase relations.
\end{abstract}
\maketitle

\section{introduction} \label{introduction}
A Weyl semimetal (WS) is a topologically nontrivial phase of matter that has been 
heavily pursued experimentally and theoretically. \cite{Wthing,weylRMP,rev1}
The band structure in these materials exhibits
Weyl points, where the conduction and valence bands intersect.
Around these points, the energy dispersion of
the quasiparticle excitations has a conical profile.
The Weyl points only appear in pairs,
which  are identified by Weyl fermions in real space, or
a source or sink of Berry curvature in momentum space. \cite{wyl5,wyl6}
This topological phase results in
the band structure in the bulk of Weyl 
semimetals being gapless, 
although  Fermi arc states 
connect pairs of Weyl points. \cite{belo}
As has been widely discussed, the presence of gapless 
surface states opens up transport channels free from
backscattering. For a conventional type-I WS, the Fermi surface is point-like,
where electron and hole states occupy
separate energy ranges above or below the Weyl point. \cite{wyl2}
By tilting the Weyl cones through the tilt parameter $\beta$, a type-II WS emerges with an open Fermi surface
containing  electron and hole pockets near the Weyl points.\cite{wyl7}
Recently, some experimental signatures of the type-II WS phase have been reported.\cite{typ2_exp2}
Also, the optical properties of type-II WSs have been investigated\cite{opt1,opt2,opt3,opt4,opt5}, including 
how tilt affects optical conductivity\cite{opt1}, absorption\cite{opt2,opt4}, and Hall conductivity.\cite{opt3}

If the Weyl semimetal is intrinsically superconducting, or becomes so through proximity effects,
it can host surface-localized 
Majorana fermions with zero energy.\cite{sc13_theor,AlidoustWS1,AlidoustBP1}
Since Majorana fermions follow non-Abelian statistics and are their own antiparticles, 
superconducting Weyl systems are anticipated to play a promising  
role in topological quantum computing. \cite{Setiawan1,Fornieri,zutic1,zutic2}
Experimentally,
superconductivity in Weyl materials has been discovered in a variety of settings, including
in $\rm {Mo Te_2}$ under pressure, and in TaAs by photoemission spectroscopy. \cite{sc1_exp,hsn4}
 It was reported that surface superconductivity can  be induced in  NbAs  through ion irradiation. \cite{bach}
Recently,  Nb-doped ${\rm Bi_2 Se_3}$ and the heavy fermion superconductor $\rm UPt_3$ 
 were  predicted to be
 Weyl superconductors. \cite{goswa}
 Apart from  naturally occurring  materials, superconductivity in Weyl semimetals 
 may also be realized through doping\cite{sc1_theor,sc4_theor,sc5_theor}. These intriguing phenomena have motivated a significant wave of interest in Weyl metal systems both theoretically and experimentally. \cite{AlidoustBP1, AlidoustBP2, AlidoustBP3, AlidoustWS1, AlidoustWS2,sc1_exp,Sirohi,Zhang,Zhang2,Chen,Xu,Fang,Bovenzi,Kononov,Kononov2,Xiao_app,shapiro2,Hou_app,Li,Das,wylsuperc_exp1,Teknowijoyo,Lu,Guguchia,Takahashi,QiaoLi,shapiro,sc2_theor,sc6_theor,sc7_theor,sc8_theor,sc9_theor,sc10_theor,sc12_theor,Gorbar,Zhou11,Sheet2,wylsuperc_exp1,Y.Xing,HWu}
 
When discussing the superconducting state of a Weyl semimetal, the
symmetry of the electron-electron coupling potential, $\Delta$,
is an important property that gives insight into the 
nature of the pairing correlations. 
Moreover, when a relationship exists
between the quasiparticle excitations in the Weyl phase
and the pair correlations,
a more complete understanding of the 
underlying electronic structure can be obtained. As a typical example,
hybrid superconducting  systems containing inhomogenous ferromagnets
can generate
odd-frequency equal-spin triplet pairs.
 In these systems, the triplet pairs
 are accompanied by a
 zero energy peak in the density of states, 
 which might be 
a signature of spin-triplet
superconducting correlations. For an overview of this phenomenon, see Refs. \onlinecite{Tanaka1,klaus_zep,zep3,G.Koren,G.Koren2,zep2} and references therein. The dominant  pairing symmetry in a system however, depends on many factors, including the band structure and strength of the pairing interactions. Thus, by implementing complimentary experimental probes that identify thermodynamic or spectroscopic signatures, the pairing symmetries and associated pairing mechanisms and types of interactions can be clarified. 

A macroscopic quantum effect that can reveal experimental signatures of the electronic properties of the Weyl semimetal is the Josephson current, which arises from a phase difference  across two superconducting segments in a junction. 
Applications involving Josephson junctions have grown due to their impact and potential 
improvements in next generation superconducting computing, nonvolatile memories, and 
magnetometers\cite{mag}, where single flux quantum circuits contain numerous 
arrangements of Josephson junctions to improve speed and sensitivity. 
To determine the thermodynamic properties of practical cryogenic Weyl devices, it is crucial to understand the behavior of charge current flow in these types of systems. 

This paper contains two main parts. Since the superconducting pair correlations can reveal hints of the host material, including lattice interactions, electronic structure, and pairing mechanisms, the symmetry of the superconducting correlations can play a significant role in 
most experimental tests. Thus, first we study the symmetry profiles of the 
superconducting correlations that a tilted WS (in the type-I and type-II phases) 
might develop through opposite-pseudospin (Pspin - the band degree of freedom) and equal-pseudospin 
phonon-mediated opposite-spin electron-electron coupling. 
To study the symmetries of the superconducting correlations, 
we construct the Green's function in the
ballistic regime. We derive analytical expressions 
for the Green's function components
and explore
 both analytically and numerically the behavior of the 
 pair correlations
for a variety of WS system parameters.
Our results show
that, depending on the parameters, a WS can support superconducting pair correlations 
with effective $s$-wave and $p$-wave symmetry classes that are even-parity and odd-parity,
consisting of opposite-Pspin and equal-Pspin channels. Importantly, we find that by properly 
tuning the parameters of a WS, including the tilting parameter $\beta$,
the effective $s$-wave and $p$-wave symmetry classes can 
evolve into $d$-wave and $f$-wave classes. Thus, our results suggest that this type of superconducting 
WS platform can serve as a ``switch" that converts the symmetry classes of the superconducting correlations. 
These predictions can be confirmed by high resolution angular point-contact spectroscopy experiments 
or Meissner-effect based experiments that are
 suitable to probe 
  the directional-dependency 
  of superconducting correlations. 
 The findings here may also stimulate further studies 
 involving 
 scenarios similar to a
 recent experiment where superconductivity 
 in a Weyl semimetal subject to external strain was studied\cite{wylsuperc_exp1}. 
 
 The second part of this paper is devoted to the transport of dissipationless charge current. 
 We make use of wavefunctions derived from an
 exact diagonalization of the
 Bogoliubov-de Gennes Hamiltonian, without applying any simplifying approximations. 
 A phase difference $\varphi$ is established across 
a Josephson junction to study
the behavior of the supercurrent in a superconductor (S)-normal metal (N)-
superconductor (S) junction. We consider the situation where superconductivity 
arises from unequal spin quasiparticles belonging to different Pspins. 
We find that by varying various parameters, including  $\beta$, and the
inversion breaking parameters of the WS,
the system can experience a  transition whereby the current 
reverses direction and changes its periodicity from $2\pi$ to $4\pi$ at the
reversal points. The inclusion of a proper inversion breaking parameter 
creates a situation where a self-biased current can flow through the junction 
with $\varphi=0$, creating a $\varphi_0$ Josephson state. 
Our results demonstrate that the self-biased supercurrent is 
controllable by different parameters of the
WS junction, including $\beta$ and junction length. 
In contrast to previous works where
an exchange field is a necessary ingredient \cite{AlidoustBP1,AlidoustBP2,zu1,zu2,zu3}, 
the $\varphi_0$ state explored here relies on inherent parameters of 
the WS Josephson junction independent of the density of nonmagnetic impurities 
and disorder that might be introduced in the system \cite{AlidoustWS2}. 
Our results in the ballistic regime are consistent 
with those explored in the diffusive limit of the quasiclassical regime\cite{AlidoustWS2}. 

In Sec.~\ref{method}, we outline the model 
Hamiltonian used and we establish the Green's function
formalism for calculating the pairing correlations that can inhabit
a Weyl semimetal with the possibility of a tilting parameter that can render 
the WS to a type-II phase. 
We also discuss the opposite-Pspin and equal-Pspin phonon-mediated electron-electron coupling 
model that is used to derive the components of 
the Green's function. We next utilize a quantum mechanical approach to find the supercurrent in 
a Josephson junction structure comprised of a WS. In Sec.~\ref{results}, we present results for both the 
pairing correlation symmetries, and the current phase relations for the charge supercurrent. Finally,
concluding remarks are given in Sec.~\ref{conclusions}.

\section{Methods} \label{method}
To model a superconducting WS system, 
supporting type-I and type-II phases \cite{sc3_theor,AlidoustWS1}, 
we use the following Bogoliubov-de Gennes Hamiltonian in spin-Nambu space:
\begin{equation} \label{Hamil}
{\cal H}({\bm k})= \left( \begin{array}{cc}
H({\bm k})-\mu\hat{1}&\hat{\Delta}\\
\hat{\Delta}^\dag& -H^{\rm T}({-\bm k})+\mu\hat{1}
\end{array}\right),
\end{equation}
in which $\mu$ is the chemical potential, ${\bm k}$ is the momentum of quasiparticles, $\hat{\Delta}$ is the superconducting gap, and $\rm T$ represents the transpose. The effective Hamiltonian $H_{\rm eff}$ in the low-energy regime, can  be then expressed as,
\begin{align}
H_\text{eff}=\int \frac{d\bm{k}}{(2\pi)^3}\check{\psi}_{\bm k}^\dag {\cal H}({\bm k}) \check{\psi}_{\bm k},
\end{align}
where the associated $1\times 8$ field operator for this Hamiltonian is given by
$\check{\psi}_{\bm k}^\dagger=[\hat{\psi_{\bm k}}^\dagger,\hat{\psi}_{-\bm k}]$, and $\hat{\psi}^\dagger_{\bm k}=( \psi_{A\ua}^\dag,\psi_{A\da}^\dag,\psi_{B\ua}^\dag,\psi_{B\da}^\dag)$,
in which the Pspin (the band degree of freedom) and spin indices are denoted by $A,B$ and $\uparrow,\downarrow$, respectively.

The  diagonal block elements of Eq.~(\ref{Hamil}) are explicitly expressed as:
\begin{align}\label{Hk}
\begin{split}
H({\bm k})= &(\beta+\gamma\tau_z )(k_z^2-Q^2)+ \eta(k_x^2+k_y^2)\tau_z +\\
&\alpha_{x} k_x\tau_x+\alpha_{y} k_y\tau_y + \alpha_z k_z\tau_z,
\end{split}
\end{align}
where the
two Weyl points are separated by $Q$  in momentum space. We will consider a broad range of orbital parameters $\alpha_{x,y,z}$,
and tilt of the Weyl cones, i.e., $\beta$, which is responsible for driving the system into the type-II phase. We will also consider both positive and negative values of the band-structure parameters $\beta$ and $\gamma$, with the latter being responsible for breaking  time reversal symmetry and band splitting. Also, a nonzero $\alpha_{z}$ shifts the Weyl nodes in energies while $\alpha_{x,y}$ can induce a gap at the nodes. The dimension of $\beta, \gamma, \eta$ is eV$\cdot{\AA}^2$ and $\alpha_{x,y,z}$ have units of eV$\cdot{\AA}$. Also, the chemical potential that enters the formalism later has units of eV and the momenta are scaled using ${\AA}$. \cite{AlidoustBP1,AlidoustBP2,AlidoustBP3,AlidoustGraph1}

In what follows, we consider a situation where the superconductivity is a 
consequence of (1) opposite-Pspin and (2) equal-Pspin phonon-mediated mechanism. 
We also consider a model of opposite-spin electron-electron interactions throughout the paper.
The first type of pairing mechanism 
is characterized by the following two-electron amplitudes: 
\begin{equation}\label{Pspin1}
\Delta_{\uparrow\downarrow}^{AB} \Big\langle\psi^\dag_{A\uparrow}\psi^\dag_{B\downarrow}\Big\rangle+\text{H.c.},
\end{equation}
where $\Delta_{\uparrow\downarrow}^{AB}$ is the gap representing BCS spin-singlet phonon mediated electron-electron coupling between the
$A$ and $B$ Pspins. The second type of pairing mechanism can be expressed by
\begin{subequations} \label{Pspin2}
\begin{align}
\Delta_{\uparrow\downarrow}^{AA} \Big\langle\psi^\dag_{A\uparrow}\psi^\dag_{A\downarrow}\Big\rangle+\text{H.c.},\\
\Delta_{\uparrow\downarrow}^{BB} \Big\langle\psi^\dag_{B\uparrow}\psi^\dag_{B\downarrow}\Big\rangle+\text{H.c.},
\end{align}
\end{subequations}
where $\Delta_{\uparrow\downarrow}^{AA}$ and $\Delta_{\uparrow\downarrow}^{BB}$ correspond to  the 
gap for BCS spin-singlet phonon mediated electron-electron coupling for each Pspin. 
In the following, to simplify notation, we suppress 
the indices so that when applicable, $\Delta_{\uparrow\downarrow}^{AB}\equiv\Delta$ 
and $\Delta_{\uparrow\downarrow}^{AA}=\Delta_{\uparrow\downarrow}^{BB}\equiv\Delta$.

We next consider the Green's functions for the Weyl semimetal system 
in the presence of superconductivity. 
The normal Green's functions $g$ and the anomalous Green's functions $f$ are defined as follows:
\begin{subequations}\label{GF_comps}
\begin{eqnarray}
&&{ g }_{\rho\rho'}^{\sigma\sigma'}(\tau,\tau'; \mathbf{ r}, \mathbf{ r}') = - \langle {\cal T}_{\tau}\psi_{\sigma\rho} (\tau,\mathbf{ r}) \psi_{\sigma'\rho'}^{\dag}(\tau',\mathbf{ r}')  \rangle,~~~~~~~\\
&&\underline{g}_{\rho\rho'}^{\sigma\sigma'}(\tau,\tau'; \mathbf{ r}, \mathbf{ r}') = - \langle {\cal T}_{\tau} \psi^{\dag}_{\sigma\rho} (\tau,\mathbf{ r}) \psi_{\sigma'\rho'}(\tau',\mathbf{ r}')  \rangle,~~~~~~~\\
&&{ f}_{\rho\rho'}^{\sigma\sigma'}(\tau,\tau'; \mathbf{ r}, \mathbf{ r}') = + \langle {\cal T}_{\tau} \psi_{\sigma\rho} (\tau,\mathbf{ r}) \psi_{\sigma'\rho'}(\tau',\mathbf{ r}')  \rangle,~~~~~~~\\
&&f_{\rho\rho'}^{\sigma\sigma'\dag}(\tau,\tau'; \mathbf{ r}, \mathbf{ r}') = +\langle {\cal T}_{\tau} \psi^{\dag}_{\sigma\rho} (\tau,\mathbf{ r}) \psi^{\dag}_{\sigma'\rho'}(\tau',\mathbf{ r}')  \rangle,~~~~~~~
\end{eqnarray}
\end{subequations}
where ${\cal T}_{\tau}$ is the time ordering operator, and $\tau, \tau'$ are the imaginary times. 
Here, $\rho,\rho'$ and $\sigma,\sigma'$ denote the spin and Pspin indices and $\langle ... \rangle$ 
denotes thermodynamic averaging.
In particle-hole space, the Green's function satisfies,
\begin{align} \label{BCSH}
\left(\begin{array}{cc}
\hat{H}(\mathbf{ r})-i\omega_n& \hat{\Delta}(\textbf{r}) \\
\hat{\Delta}^*(\textbf{r}) &  {{\cal T}_\text{t}}{\hat{H}(\mathbf{ r})}{{\cal T}_\text{t}}^\dag+i\omega_n
\end{array}\right)
\check{g}(i\omega_n;\mathbf{r},\mathbf{r}')
=\delta(\mathbf{r}-\mathbf{r}'),
\end{align}
in which $\omega_n=\pi (2n+1) k_B T $ is the Matsubara frequency, $n\in { Z}$, $k_B$ is the Boltzman constant, $ T $ is temperature, 
${\cal T}_\text{t}$ is the time reversal operator, and $\hat{\Delta}(\mathbf{ r})$ is the superconducting gap in real space. 
Note that the Hamiltonian in real space $\hat{H}(\mathbf{ r})$ is obtained by 
replacing $i{\bm k}\equiv (\partial_x,\partial_y,\partial_z)$ in $H({\bm k})$, in Eq.~(\ref{Hk}). 
The matrix form of the Green's function reads:
\begin{equation}
\check{g}(i\omega_n;\mathbf{ r},\mathbf{ r}')=\left(  \begin{array}{cc}
\hat{g}(i\omega_n;\mathbf{ r},\mathbf{ r}') & \hat{f}(i\omega_n;\mathbf{ r},\mathbf{ r}')\\
\hat{f}^\dag(i\omega_n;\mathbf{ r},\mathbf{ r}') & \hat{\underline{g}}(i\omega_n;\mathbf{ r},\mathbf{ r}')
\end{array}  \right),
\end{equation}
where we denote the
$4\times 4$ matrices by the symbol, $\hat{...}$ , and $8\times 8$ matrices by the  symbol, $\check{...}$ .

\begin{figure*}[thp]
\includegraphics[width=17.0cm,height=7.80cm]{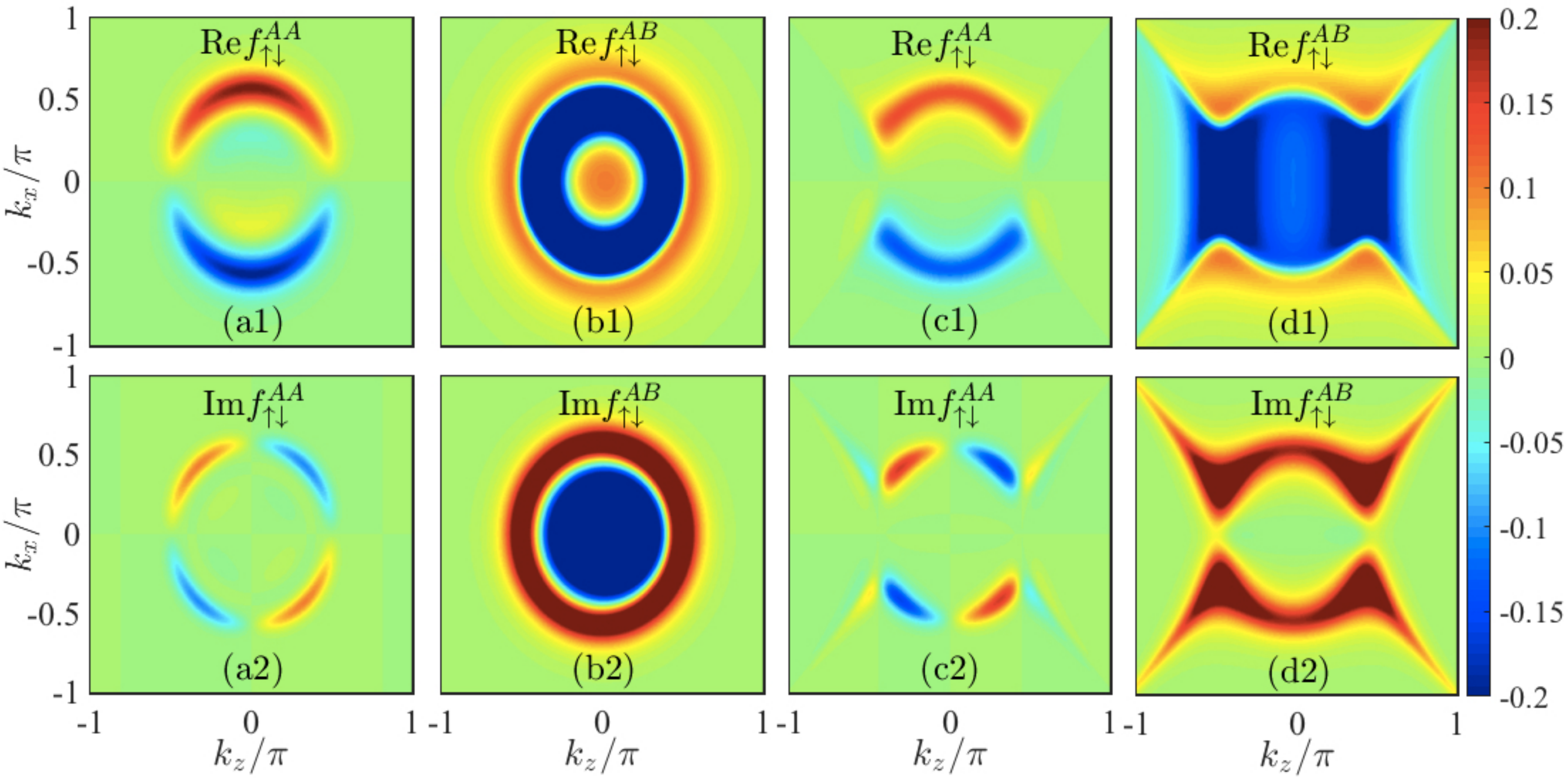} 
\caption{ (Color online). Real and imaginary  parts of the
superconducting correlations for the \textit{opposite}-pseudospin phonon mediated electron-electron interaction
plotted as a function of momenta $k_x$ and $k_z$.
For illustrative purposes, the momentum component 
 $k_y$ is set to zero.
The following pair correlations are shown:
$f_{\uparrow\downarrow}^{AA}$ and $f_{\uparrow\downarrow}^{AB}$. 
(a)-(b) ($\gamma>\beta $) We use the representative parameters
 $\gamma=1.3$, $\beta = 0.1$, $\alpha_z=\alpha_{xy}=0.2$, 
 $\mu = 0.5$, $\eta = 1$, and $Q=0.4\pi$. (c)-(d) ($\gamma<\beta $) We use the representative parameters
 $\gamma=0.1$, $\beta = 1.3$, $\alpha_z=\alpha_{xy}=0.2$, 
 $\mu = 0.5$, $\eta = 1$, and $Q=0.4\pi$.  }
 \label{fig1}
\end{figure*}

To calculate the supercurrent flowing through a Weyl semimetal Josephson junction, 
we assume a steady state situation where the
charge density $\rho$ in the system is constant in time, namely, $\partial\rho/\partial t \equiv 0$.
To ultimately calculate the charge current,
we employ the  
quantum mechanical definition of the time-evolution 
of $\rho$
for a generic system that ensures charge conservation
in the steady state, namely,
\begin{equation}\label{crntdif}
\begin{split}
\frac{\partial \rho}{\partial t}=\lim\limits_{{\bm r}\rightarrow {\bm r}'}\sum\limits_{\sigma\rho\nu\sigma'\rho'\nu'}\frac{1}{i\hbar}\Big[ \psi^\dag_{\sigma\rho\nu}({\bm r}'){\cal H}_{\sigma\rho\nu\sigma'\rho'\nu'}({\bm r})\psi_{\sigma'\rho'\nu'}({\bm r})\\-\psi^\dag_{\sigma\rho\nu}({\bm r}'){\cal H}_{\sigma\rho\nu\sigma'\rho'\nu'}^\dag({\bm r}')\psi_{\sigma'\rho'\nu'}({\bm r})\Big].
\end{split}
\end{equation}
Here ${\cal H}_{\sigma\rho\nu\sigma'\rho'\nu'}$ is the component form of Eq.~(\ref{Hamil}) 
with the Pspin, spin, and particle-hole ($\nu,\nu'$) indices
written explicitly. 
We now use the law of current conservation to arrive at the following 
quantum mechanical expression for the current density:
\begin{align}
{\bm J} =\int \hspace{-.1cm} d{\bm r}\{\hat{\psi}^\dagger({\bm r}) \overrightarrow{{\cal H}}({\bm r})\hat{\psi}({\bm r})-
\hat{\psi}^\dagger({\bm r}) \overleftarrow{{\cal H}}({\bm r})\hat{\psi}({\bm r}) \},
\end{align}
where ${\cal H}({\bm r})$ is given by Eq.~(\ref{Hamil}), after the substitution $i{\bm k}\equiv (\partial_x,\partial_y,\partial_z)$. The arrows point to the specific wavefunctions that are acted on by the derivative operators in the Hamiltonian. Note that any property of the system (including anisotropy and inhomogeneity) are indeed imbedded within the Hamiltonian of the system, associated wavefunctions, and boundary conditions (see below for more details).  

\section{results and discussions} \label{results}
In this section we present results that depict representative behavior of the pair 
correlation profiles that may evolve in Weyl semimetal systems.
We examine the momentum-space profiles and show that
by varying  experimentally relevant parameters,  
the effective symmetry classes of the
superconducting correlations can 
undergo significant transformations.
To compliment this,
the charge supercurrent flow is then studied 
in a ballistic SNS Josephson  structure, where  a phase difference 
between the two superconducting electrodes is established.
The direction and periodicity of supercurrent flow,
and the possible emergence of $0$-$\pi$ transitions are investigated
as a function of orbital effects, as well as
tilt and spacing of the 
Weyl cones.

\subsection{Symmetry profile of superconducting pair correlations}
Here we present the anomalous Green's function 
$\hat{f}(i\omega;\bm{k})$ and explore different 
symmetries that may arise by manipulating the 
Weyl semimetal material parameters.
In the calculations that follow, we consider a representative node separation corresponding to $Q=0.4\pi$, and note that our conclusions are independent of this specific choice. 
For simplicity, we consider the zeroth mode of the Matsubara frequency, following the fact that parity is not affected within the frequency domain.
Each component of $\hat{f}(i\omega;\bm{k})$ describes a specific pairing correlation function. To further simplify our calculations,
we consider a system that is invariant in all directions so the
momenta $k_{x,y,z}$ are 
good quantum numbers and can be treated algebraically in momentum space. 
By solving Eq.~(\ref{BCSH}) in conjunction with the opposite-Pspin phonon mediated coupling of
Eq. (\ref{Pspin1}), we end up with the following expressions for the superconducting correlations:
\begin{subequations} \label{fab_all}
\begin{align}
f_{\uparrow\uparrow}^{AA}=&0\\
  f_{\uparrow\downarrow}^{AA} =& \frac{2}{{\cal D}} \alpha_{xy} \Delta \underbrace{(k_x - i k_y)}_{\cal P_-} \underbrace{(\Lambda + \mu + \Omega(\gamma - \beta))}_{\cal D_-},\label{faa_ud}\\
 f_{\uparrow\uparrow}^{AB}=&0\\ \nonumber
  f_{\uparrow\downarrow}^{AB} = &-\frac{\Delta}{\cal D}
 \left[
 -(\Omega( \gamma -\beta)  +\Lambda +i
   \omega_n) \right.\\&(\Omega  (\beta +\gamma )+\Lambda +i
  \omega_n)+\Delta ^2+\label{fab_ud}\\\nonumber &\left.\alpha_{xy}^2
   \left(k_x^2+k_y^2\right)+\alpha_z^2
   k_z^2+2 \alpha_z k_z (\mu -\beta  \Omega)-2\mu\beta  \Omega
\right],
\end{align}
\end{subequations}
in which ${\cal D}$ is given by: 
\onecolumngrid
\begin{align} \label{bigN}
\nonumber {\cal D}=&(k_x^2 + k_y^2) \alpha_{xy}^2 (-\mu^2 + (k_x^2 + 
       k_y^2) \alpha_{xy}^2 + \Delta^2 + 
    2 \mu (\beta \Omega - 
       i\omega_n) + (k_z \alpha_{z} + \Lambda + (\beta + 
\gamma) \Omega - 
       i \omega_n) (k_z \alpha_{z} + \Lambda - \beta 
\Omega + \gamma \Omega + 
       i \omega_n)) + \\&\Delta^2 ((k_x^2 + 
       k_y^2) \alpha_{xy}^2 + \Delta^2 - (\mu + 
       k_z \alpha_{z} + \Lambda - \beta \Omega + 
\gamma \Omega + i \omega_n) (-\mu - 
       k_z \alpha_{z} + \Lambda + (\beta + \gamma) 
\Omega + i \omega_n)) + \nonumber \\&(\Delta^2 (-\mu + 
       k_z \alpha_{z} + \Lambda + (\beta + \gamma) 
\Omega - 
       i \omega_n) + (\mu^2 - (k_x^2 + 
          k_y^2) \alpha_{xy}^2 - (k_z \alpha_{z} + \Lambda + (
\beta + \gamma) \Omega - 
          i \omega_n) (k_z \alpha_{z} + \Lambda - \beta 
\Omega + \gamma \Omega + i \omega_n) +\nonumber \\& 
       \mu (-2 \beta \Omega + 2 i \omega_n)) (-\mu - 
       k_z \alpha_{z} + \Lambda + (\beta + \gamma) 
\Omega + i \omega_n)) (-\mu + 
    k_z \alpha_{z} - \Lambda + \beta \Omega - 
\gamma \Omega + i \omega_n)
\end{align}
\twocolumngrid \noindent
where we have defined $\Omega = k_z^2 - Q^2$, and $\Lambda = \eta (k_x^2 + k_y^2 )$.
In a similar fashion, and using the
symmetry relations appropriate for the Green's function components,
one can obtain the other components of $\hat{f}(i\omega;\bm{k})$. 
We have 
derived all components of $\check{g}(i\omega;\bm{k})$, however,  
for the purposes shown in this paper, the components presented above are adequate.

\begin{figure*}[thp]
\includegraphics[width=17.0cm,height=7.80cm]{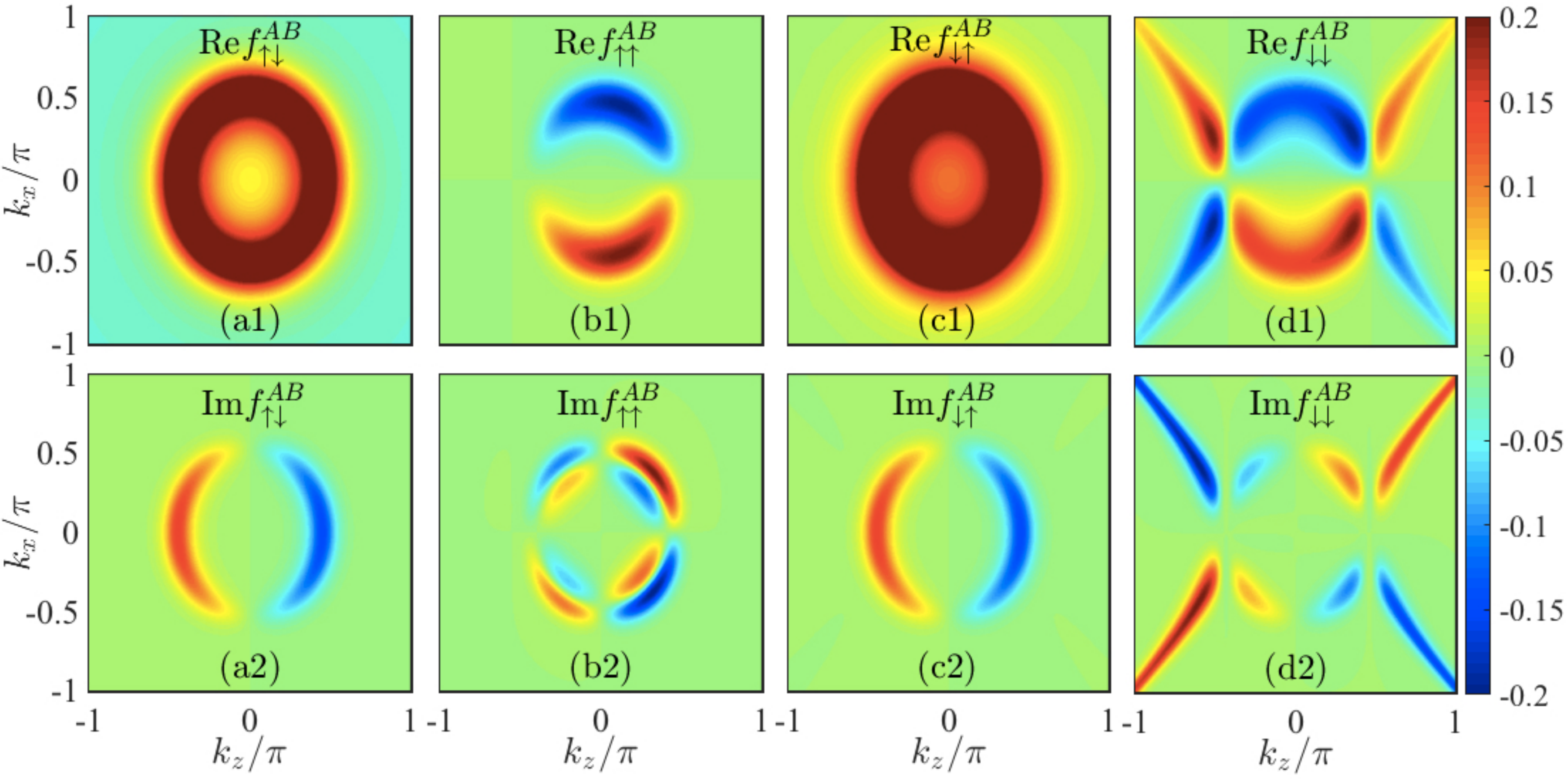} 
\caption{ (Color online). Real and imaginary  parts of the
superconducting correlations for the \textit{equal}-pseudospin phonon mediated electron-electron interaction
plotted as a function of momenta $k_x$ and $k_z$.
The following pair correlations are shown:
$f_{\uparrow\downarrow}^{AA}$ and $f_{\uparrow\downarrow}^{AB}$. 
The parameters here are the same as in Fig.~\ref{fig1}.
 }
 \label{fig2}
\end{figure*}

Here, we have set $\alpha_x=\alpha_y=\alpha_{xy}$ to further simplify our
expressions. Note that 
if we set $\alpha_{xy}=0$, Eq.~(\ref{bigN}) reduces to,
\begin{align}\label{N_reduced}
{\cal D} = &-\Delta^2 + (-\mu + 
    k_z \alpha_z - \Lambda+ \beta \Omega - 
\gamma \Omega + i \omega_n) \\&(\mu - 
    k_z \alpha_z - \Lambda- (\beta + \gamma) 
\Omega + i \omega_n),
\end{align}
and we also  find,
\begin{subequations} \label{fab_all3}
\begin{align}
f_{\uparrow\uparrow}^{AA}=&0,\\
  f_{\uparrow\downarrow}^{AA} = &0,\\
  f_{\uparrow\uparrow}^{AB}= &0,\\
 f_{\uparrow\downarrow}^{AB} =&\frac{\Delta}{\cal D}.
 \end{align}
\end{subequations}
As seen, the orbital term is indeed responsible for 
inducing equal-Pspin pair correlations in WS 
so $f_{\uparrow\downarrow}^{AA}$is directly proportional to $\alpha_{x,y}$.  
The expression governing the equal-Pspin 
component of the anomalous Green's 
function (when $\alpha_{x,y}\neq 0$) in Eq.~(\ref{faa_ud}) has several important features:
The numerator is proportional to ${\cal P}_{-}=k_x- ik_y$, 
which is of odd-parity, and corresponds to an effective $p$-wave  symmetry 
($p_x- i p_y$). 
The second term in parentheses, ${\cal D}_{-}$, containing $\Omega$ and $\Lambda$,
is of even-parity and has a symmetry expressible in the form $|a|(k_x^2+k_y^2)\pm |b| k_z^2$, where $a$ and $b$ are coefficients
that depend on  $\eta, \gamma$, and $\beta$. Specifically, from ${\cal D}_+=\Gamma+\mu +\Omega(\gamma+\beta)$, we find that $a\equiv \eta$ and $b\equiv (\gamma-\beta)$.
Hence, depending on the choice of parameters (for example, in the regime where $\beta>\gamma$ or $\beta<\gamma$), the 
symmetry classification for the second term can be of the $s$-wave  
($|a|(k_x^2+k_y^2)+|b| k_z^2$) or $d$-wave ($|a|(k_x^2+k_y^2)- |b| k_z^2$) types. 
Depending on the symmetry of ${\cal D}_{-}$, the numerator of 
Eq.~(\ref{faa_ud}) belongs to the $p$-wave or $f$-wave symmetry classes.
The complete symmetry profiles of $f_{\downarrow\uparrow}^{AA,BB}$ also depend on
 the  symmetry of the
denominator ${\cal D}$.
Therefore, in what follows, we study {\it effective} symmetries of the
superconducting pair correlations. Upon variations  of the 
parameters $\eta, \gamma$, and $\beta$ in Eqs.~(\ref{faa_ud}) and (\ref{fab_ud}), it is evident that
the cumulative effect
on  $f_{\uparrow\downarrow}^{AA}$ and $f_{\uparrow\downarrow}^{AB}$   
is to produce a
spatial symmetry 
that can be classified as effectively $p$-wave and $f$-wave. 
Although  ${\cal D}$ can have an effective $s$-wave symmetry, depending on $\eta, \gamma,\beta$, if $\alpha_z$ is nonzero, the
overall  symmetry of $f_{\uparrow\downarrow}^{AA}$ and $f_{\uparrow\downarrow}^{AB}$
can
change as seen in
Eqs.~(\ref{bigN}) and (\ref{N_reduced}). 
Although more lengthy, an analysis of the components 
$f_{\uparrow\downarrow}^{BB}$, and $f_{\uparrow\downarrow}^{BA}$ reveals similar possibilities. 
For instance, if we set $\alpha_z=0$, we see that the 
numerators of $f_{\uparrow\downarrow}^{BA}$ can show  either $s$-wave or $d$-wave effective
symmetries, 
depending again, on the parameters $\eta, \gamma$, and $\beta$. 
Thus, 
for the phonon mediated unequal-Pspin scenario, 
the superconducting pairing correlations  
can change their effective spatial symmetry profiles from one symmetry class to another, simply by manipulating the material
parameters of a Weyl semimetal. 
As recently demonstrated, control
of these parameters is possible by, for example, applying 
strain to the system.\cite{AlidoustBP1,AlidoustBP2,AlidoustBP3,AlidoustGraph1} Also, unconventional pairings and their symmetries were recently studied in systems with and without the lack of inversion symmetry. \cite{yuri1,yuri2,yuri3} It is also worth noting that our results above illustrate that superconducting Weyl semimetals can develop odd-frequency symmetries. \cite{HaltFSFH,zep2,AlidoustGraph1,AlidoustBP3,AlidoustWS2}

\begin{figure*}[thp]
\includegraphics[width=17.0cm,height=7.80cm]{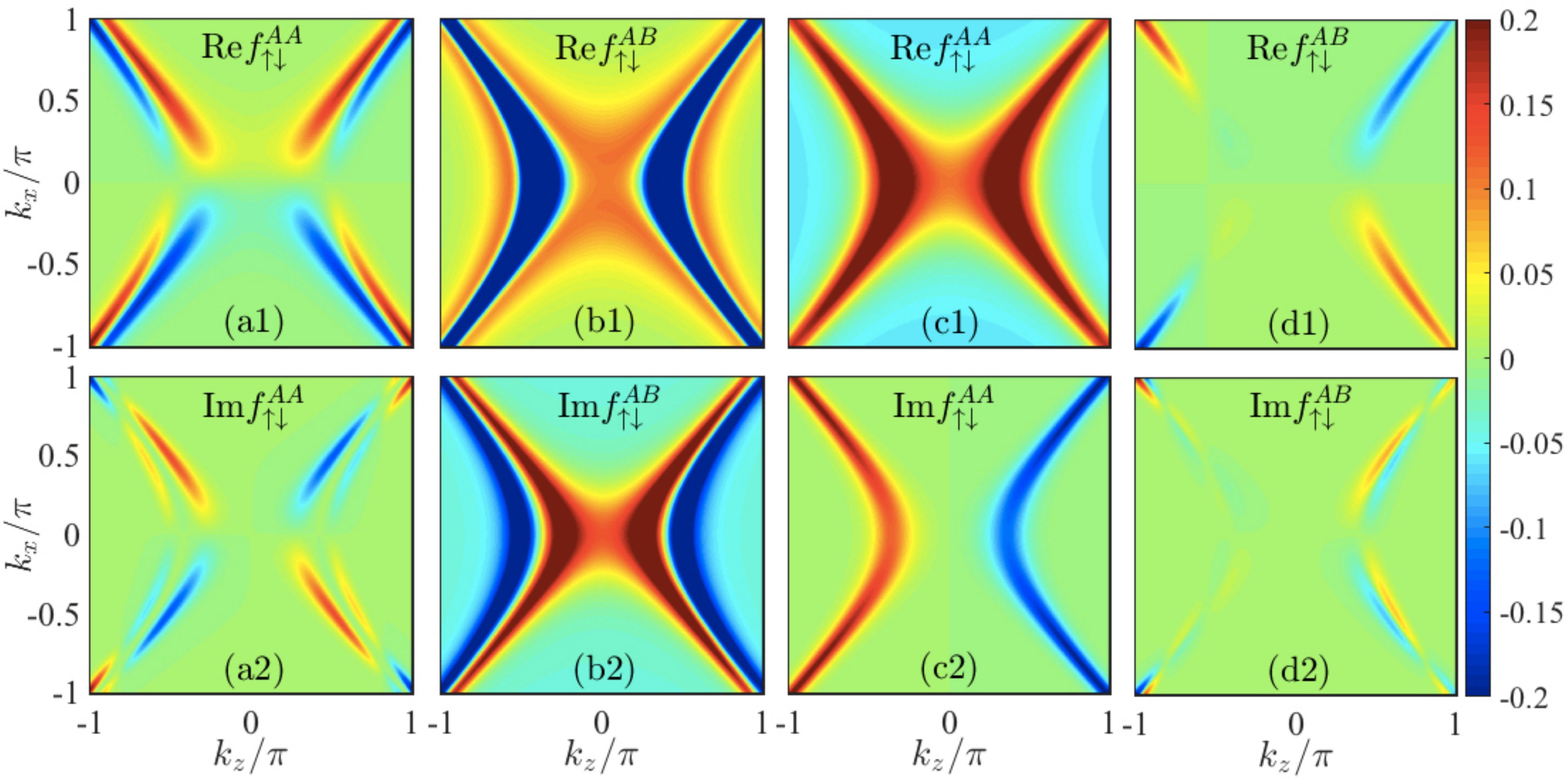} 
\caption{ (Color online). Real and imaginary parts of the
superconducting correlations $f_{\uparrow\downarrow}^{AA}$ and $f_{\uparrow\downarrow}^{AB}$ 
plotted as a function of momenta $k_x$ and $k_z$. (a)-(b) The \textit{opposite}-pseudospin phonon mediated 
electron-electron interaction is considered and parameters are identical to those of Figs.~\ref{fig1}(a)-\ref{fig1}(b) 
except we now set $\gamma=-1.3$. (c)-(d) The \textit{equal}-pseudospin phonon mediated electron-electron 
interaction is set and parameters are identical to those of Figs.~\ref{fig2}(a)-\ref{fig2}(b),
except now we flip the sign of $\gamma$ to minus $\gamma=-1.3$. 
 }
 \label{fig2.5}
\end{figure*}

To further shed light on the symmetry profiles of 
these superconducting correlations,
in Figs.~\ref{fig1}-\ref{fig2.5}, we present a few representative profiles 
for the real and imaginary parts of the equal-Pspin and opposite-Pspin components 
as a function of the 
momenta $k_x$ and $k_y$. 
The upper row of panels corresponds to the real parts, while the bottom 
row is for the imaginary parts. For simplicity, we consider zero momentum along $y$, $k_y=0$. 
We also consider here two regimes ($\beta<\gamma$ and $\beta>\gamma$) and equal strengths of 
the orbital parameters:  $\alpha_z=\alpha_{xy}=0.2$. 
In Figs.~\ref{fig1}(a)-\ref{fig1}(b) we implement a representative 
set of parameters: $\gamma=1.3$,  $\beta=0.1$, $\mu = 0.5$, $\eta = 1$, and $Q=0.4\pi$. As seen in Figs. \ref{fig1}(b1) and \ref{fig1}(b2), the {\it effective} 
symmetry of the $f_{\uparrow\downarrow}^{AB}$ correlations
are of the $s$-wave 
type, which is indeed fully consistent with our discussion
of Eq.~(\ref{fab_ud}). 
Likewise, Figs.~\ref{fig1}(a1) and \ref{fig1}(a2) show that the effective symmetry of both imaginary and real parts of $f_{\uparrow\downarrow}^{AA}$ are of the $p$-wave type, consistent 
with Eqs.~(\ref{faa_ud}). We note that here we 
consider the regime where $\gamma>\beta$.

To illustrate how the effective $d$-wave 
and $f$-wave symmetries might evolve, we now set $\gamma=0.1$ and $\beta=1.3$, 
corresponding to the $\gamma<\beta$ regime. 
Note that, according to the discussion above, one is able to achieve similar symmetry changes in the effective symmetry profiles (described below) 
by tuning and/or reversing the sign of the parameters $\gamma$, $\eta$, and $\beta$.
The resultant 
superconducting correlations are shown  in Figs.~\ref{fig1}(c)-\ref{fig1}(d). 
We clearly observe in Figs.~\ref{fig1}(b1), \ref{fig1}(b2), \ref{fig1}(d1), and \ref{fig1}(d2) 
that the previous effective $s$-wave symmetry has transformed into an effective 
$d$-wave class. 
Likewise, 
Figs.~\ref{fig1}(a1), \ref{fig1}(a2), \ref{fig1}(c1), and \ref{fig1}(c2) 
illustrate that  the previous effective $p$-wave symmetry has become  of the $f$-wave type, 
which arises from the multiplication of terms with effective 
$p$-wave and $d$-wave symmetries, as  discussed in 
connection with Eqs.~(\ref{faa_ud}) and (\ref{fab_ud}). 
As for the effects of increasing the tilt parameter $\beta$, 
Eqs.~(\ref{fab_all}) reveal that there 
are two competing effects that determine the
final symmetry of the pair correlations:
As $\beta$ increases, the pair correlations involving the term $(\gamma-\beta)$
decrease their $s$-wave character [$|a|(k_x^2+k_y^2)+|b| k_z^2$],
which is countered by 
the parameter $\gamma$ that drives the correlations into a $d$-wave symmetry class [$|a|(k_x^2+k_y^2)-|b| k_z^2$].
Thus, a multitude of possibilities arise for the pair correlation symmetries 
depending on the material properties and whether the Weyl semimetal is type-I or type-II.

To gain additional insight, we have also calculated the superconducting pairing correlations 
starting from an equal-Pspin scenario given by Eqs.~(\ref{Pspin2}). After performing the 
calculations, we find the following expressions for the components of anomalous Green's function: 
\begin{subequations} \label{EPspin}
\begin{align}
f_{\uparrow\uparrow}^{AA}=&0,\\
  f_{\uparrow\downarrow}^{AA} = &\frac{1}{\cal D}[ \Delta ((k_x - 
      i k_y)^2 \alpha_{xy}^2 + \Delta^2 + (\mu + 
\Lambda+\nonumber\\& (-\beta + \gamma) \Omega)^2 - (k_z 
\alpha_{z} + i \omega_n)^2) ],\label{fud_aa}\\
  f_{\uparrow\uparrow}^{AB}= &0,\\
 f_{\uparrow\downarrow}^{AB} =&-\frac{1}{\cal D}[  2 \alpha_{xy} \Delta (k_x (\mu + 
      k_z \alpha_{z} - \beta \Omega) + 
   i k_y (\Lambda+ \gamma \Omega + i \omega_n)) ],\label{fud_ab}
 \end{align}
\end{subequations}
\onecolumngrid
\begin{align} \label{bigD2}
\nonumber {\cal D}=&-(-k_x + i k_y) \alpha_{xy} ((k_x - 
       ik_y) \alpha_{xy} \Delta^2 + (k_x + 
       ik_y) \alpha_{xy} (-\mu^2 + (k_x^2 + k_y^2) \alpha_{xy}^2 + 
       2 \mu (\beta \Omega - 
          i \omega_n) + (k_z \alpha_{z} + \Lambda+ (\beta 
+ \gamma) \Omega - 
          i \omega_n) \nonumber\\&(k_z \alpha_{z} + \Lambda- \beta 
\Omega + \gamma \Omega + 
          i \omega_n))) + \Delta^2 ((k_x + 
       i k_y)^2 \alpha_{xy}^2 + \Delta^2 + (-\mu + 
\Lambda+ (\beta + \gamma) \Omega)^2 - (k_z 
\alpha_{z} - 
      i \omega_n)^2) + \nonumber\\&(-\Delta^2 (\mu + 
       k_z \alpha_{z} + \Lambda- \beta \Omega + 
\gamma \Omega + 
       i\omega_n) + (\mu^2 - (k_x^2 + 
          k_y^2) \alpha_{xy}^2 - (k_z \alpha_{z} + \Lambda+ (
\beta + \gamma) \Omega - 
          i \omega_n) \nonumber\\&(k_z \alpha_{z} + \Lambda- \beta 
\Omega + \gamma \Omega + i \omega_n) + 
       \mu (-2 \beta \Omega + 2 i \omega_n)) (-\mu - 
       k_z \alpha_{z} + \Lambda+ (\beta + \gamma) 
\Omega + i \omega_n)) (-\mu + 
    k_z \alpha_{z} - \Lambda+ \beta \Omega - 
\gamma \Omega + i \omega_n).
\end{align}
\twocolumngrid \noindent
We see that this kind of pairing interaction results in complicated expressions 
for the superconducting correlations, specifically for the denominator.  
Therefore, we evaluate them numerically in Fig.~\ref{fig2}. Similar to the previous case, 
if we set $\alpha_{xy}=0$, the expressions simplify considerably:
\begin{subequations} \label{}
\begin{align}
f_{\uparrow\uparrow}^{AA}=&0,\\
  f_{\uparrow\downarrow}^{AA} = &\frac{\Delta}{\cal D},\\
  f_{\uparrow\uparrow}^{AB}= &0,\\
 f_{\uparrow\downarrow}^{AB} =&0 ,
 \end{align}
\end{subequations}
\begin{equation}
{\cal D} =\Delta^2 + (-\mu + \Lambda + (\beta + \gamma) 
\Omega)^2 - (k_z \alpha_z - i \omega_n)^2.
 \end{equation}
In agreement with our initial assumption for the electron-electron interactions [i.e., Eqs.~(\ref{Pspin2})], 
we find that $f_{\uparrow\downarrow}^{AA}$ (or equivalently $f_{\uparrow\downarrow}^{BB}$) is the
 only nonzero component with  even-parity when $\alpha_z=0$, which is evidently of the $s$-wave type. 
 In Fig.~\ref{fig2}, we have plotted the equal-Pspin and opposite-Pspin 
 components of Green's function as a function of $k_x, k_z$, where $k_y=0$. 
 The parameters set are identical to that used in Fig.~\ref{fig1}. 
 As seen in Figs. \ref{fig2}(a)-\ref{fig2}(b) and  \ref{fig2}(c)-\ref{fig2}(d), 
 when we move from the $\gamma>\beta$ regime to the regime where $\gamma<\beta$, 
 only $f_{\uparrow\downarrow}^{AB}$ shows a discernible change 
 in its spatial symmetry,
 while $f_{\uparrow\downarrow}^{AA}$ remains almost entirely intact. 
 This is in contrast to the previous cases where both components 
 $f_{\uparrow\downarrow}^{AB}$ and $f_{\uparrow\downarrow}^{AA}$ 
 show changes in their symmetries. Nevertheless, we can conclude 
 that the symmetry change is robust and can occur  for the two electron-electron 
 interaction scenarios considered in this paper
 (see also the following discussions).

Figure~\ref{fig2.5} illustrates the pairing correlations 
$f_{\uparrow\downarrow}^{AB}$ and $f_{\uparrow\downarrow}^{AA}$,
where we have simply flipped the sign of $\gamma=-1.3$, and kept $\beta=0.1$ unchanged. 
In Figs.~\ref{fig2.5}(a)-\ref{fig2.5}(b), we plot the correlations for when the interaction between 
electrons with opposite spins are of the opposite-Pspins [Eq.~(\ref{Pspin1})].
In contrast,
Figs.~\ref{fig2.5}(c)-\ref{fig2.5}(d) correspond to  equal-Pspin 
opposite-spin electron-electron interactions. It can be seen that by 
flipping the sign of $\gamma$, one can 
achieve pronounced symmetry changes in the superconducting correlations. 
The same feature can be observed when the sign of $\beta$ is reversed in the opposite 
regime where $\beta>\gamma$ (not shown).  

Note that the possibility of symmetry change in spatial profiles of pair correlations 
relies on a sign change of the momenta coefficients. Such situations can be precisely determined 
by  first-principles calculations; e.g., when the system experiences external strain or is subject to 
doping and intercalation. A practical example is
when a black phosphorus monolayer is under the application 
of an external strain, as was recently discussed \cite{AlidoustBP1,AlidoustBP2,AlidoustBP3,AlidoustGraph1}. 
The band-structure parameters  were obtained using density-functional-theory computations,
and symmetry calculations demonstrated that the momenta coefficients can 
change signs depending on the strength and direction of the 
applied in-plane strain. 
Hence, angular point-contact tunneling spectroscopy experiments 
or Meissner effect imaging are ideal 
probes to confirm the symmetry-change predictions discussed above.  

An important quantity that can be additionally accounted for in our calculations and  serve as a 
controlling parameter, is an exchange field \cite{AlidoustBP1,AlidoustBP2}. Although we have not explicitly written it in our formulations, to incorporate an exchange field with three nonzero components $h_i$, the terms $h_x\rho_x+h_y\rho_y+h_z\rho_z$ should be attached to each Pspin [see Refs. \onlinecite{AlidoustBP1,AlidoustBP2,AlidoustBP3,AlidoustGraph1}], in which $\rho_i$ are Pauli matrices in spin space. We have investigated the superconducting pair correlations when the exchange field is nonzero in different directions. Our results (not shown) illustrate that the change of symmetry classes predicted above occurs for the spin-triplet superconducting correlations as well in the presence of an exchange field.

Our theoretical methods may also be extended to a scenario similar to
the one investigated  in a recent experiment that observed controlled 
superconductivity in $\rm MoTe_2$ by the application of an external mechanical strain \cite{wylsuperc_exp1}. 
It was observed that by the application of strain to $\rm MoTe_2$, the superconducting critical temperature $T_c$  
increased. 
In consideration of this,
it would be of interest to 
study
$T_c$  for 
WS systems over
a broad range of material and geometrical parameters.
In this case,
 the superconducting gap function 
 should be solved self-consistently 
to extract $T_c$, which is
however, a nontrivial and formidable task that 
is postponed for future works. \cite{HaltSFH,HaltFSFH} 
  
\begin{figure}[t!]
\includegraphics[clip, trim=2cm 7.cm 5cm 4cm, width=8.50cm,height=4.0cm]{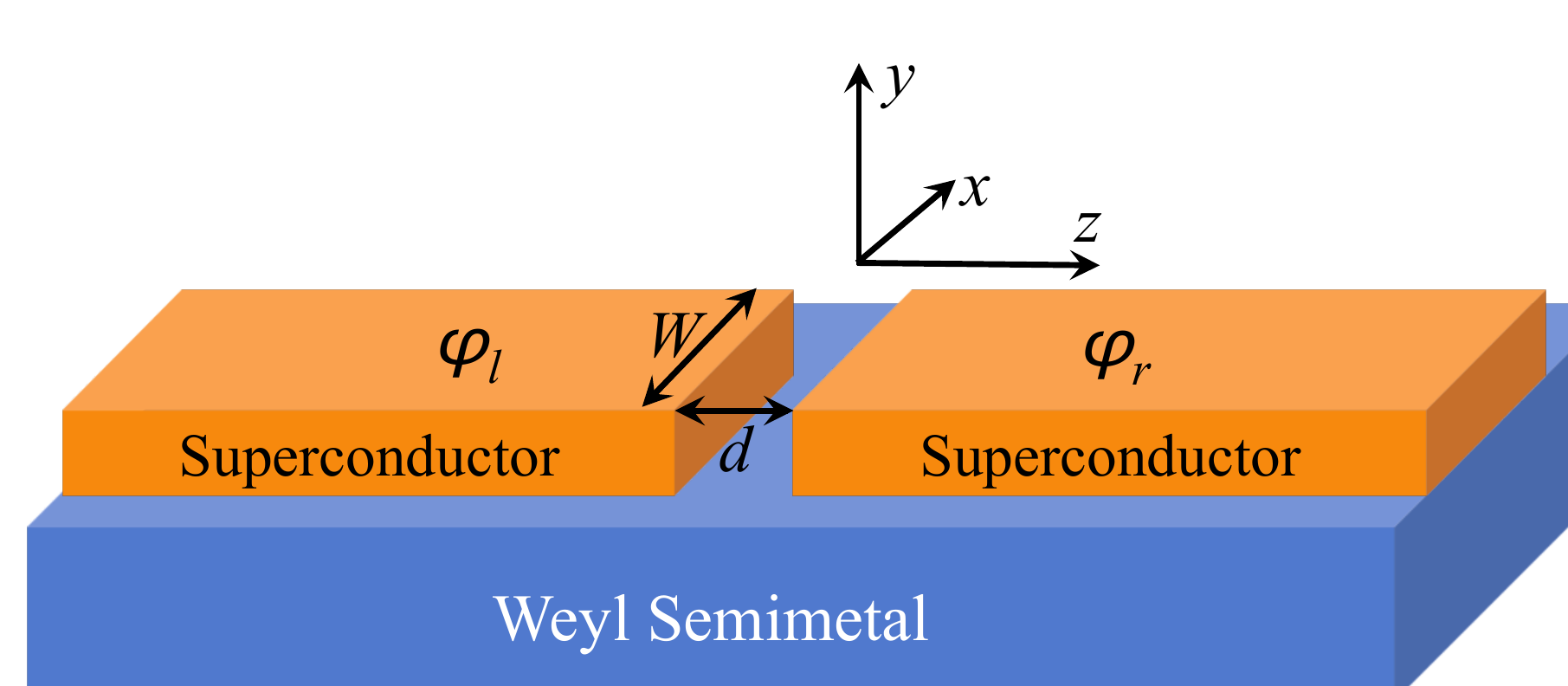}
\caption{ (Color online). 
Schematic of a Weyl semimetal Josephson junction. 
The junction interfaces lie in
the $xy$ plane, with the normal to the interfaces 
along the $z$ direction. The interfaces 
between the normal state  Weyl semimetal and 
the superconductors are  located at $z=\pm d/2$. 
The width of the junction is $W$ and the 
macroscopic phase of the left and right superconductors are labeled 
by $\varphi_l$ and $\varphi_r$, respectively. }
\label{fig3}
\end{figure}

\begin{figure*}[t]
\includegraphics[width=18.0cm,height=6.50cm]{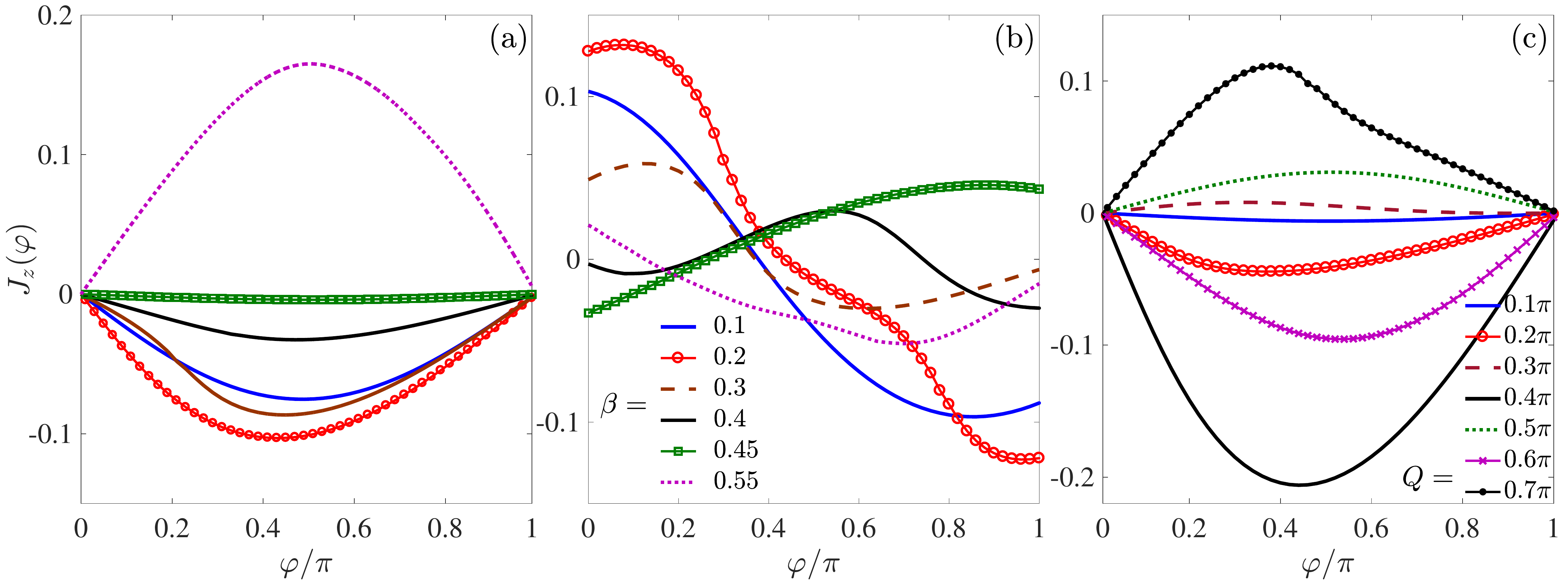} 
\caption{ (Color online).
Normalized current flowing in the $z$ direction, $J_z(\varphi)$, as a 
function of the superconducting phase difference $\varphi$ applied 
across the junction for various values of  the tilting parameter $\beta$ 
and separation of Weyl nodes $Q$ (see legends). 
In (a) the orbital term $\alpha_z$ is set zero. 
In (b) the orbital term $\alpha_z$ possesses a representative finite value of $0.2$. 
In (c) we set $\alpha_z$ zero and vary $Q$. 
The other parameters are set as follows: $\gamma=0.5$, $\beta=0.2$, $\alpha_{x,y}=1$,
and  $\eta=1$. }
\label{fig4}
\end{figure*}

\subsection{Charge Supercurrent}  \label{current}
We now proceed to investigate  the supercurrent flow
through a Weyl
Josephson junction.
We consider a SNS type structure with a phase difference $\varphi=\varphi_l-\varphi_r$
established between the two superconducting electrodes  [see Fig.~\ref{fig3}]. To compute the total charge current, we take the 
current density perpendicular to the interface, which is in our case $j_z$, and  
integrate over the $x$ direction,
 i.e., $I(\varphi)=\int_0^{W_x}  dx j_z(x,z,\varphi)$. 
Note that the current is independent of the $z$-coordinate 
as required by charge conservation. 
Upon diagonalizing ${\cal H}({\bm r})$, we calculate the
spinors $\psi_{\sigma\rho\nu}$ 
within each region. Next, To obtain the appropriate wavefunction in the 
normal region of the Josephson configuration, 
we match wavefunctions at the left $\hat{\psi}_l(z=-d/2)=\hat{\psi}_r(z=-d/2)$ 
and right boundaries $\hat{\psi}_r(z=+d/2)=\hat{\psi}_l(z=+d/2)$, and apply the continuity condition 
$\partial_{\bm k} {\cal H}_l({\bm k})\hat{\psi}_l=\partial_{\bm k} {\cal H}_r({\bm k})\hat{\psi}_r$, 
at the left and right superconductor interfaces. At the left boundary, ${\cal H}_l$ and ${\cal H}_r$ 
correspond to the Hamiltonians of the superconducting and normal regions ($\hat{\psi}_l$ and $\hat{\psi}_r$ 
are their associated wavefunctions), respectively, whereas at the right boundary 
they represent the Hamiltonians
of the normal and superconducting regions, respectively. 
We note that, in order to accommodate a wide range of parameters without 
missing the important physics, we make no assumptions and approximations regarding the 
strength of the parameters involved that would normally simplify the  wavefunctions. 
For example, the
quasiclassical approximation assumes that $\mu$ is the largest energy in the system, 
which considerably  simplifies the wavefunctions, but at the cost of losing 
important information for regimes where $\mu$ is small \cite{AlidoustWS1,AlidoustWS2,AlidoustBP1,AlidoustBP2}. 
The resultant analytic expressions
involving $1\times 8$ spinors are therefore very cumbersome, and  we omit the explicit expressions, 
presenting  numerical results only when evaluating observable quantities.  
The junction length is fixed at a finite representative value ($d\approx 7.5$ nm), and we assume that the junction width $W$ 
is sufficiently large ($W\approx 100$ nm) so that  boundary effects along the width direction are negligible. 
The central junction segment 
is undoped with $\mu_N=0$, 
while the chemical potential in the 
superconducting segments have the representative value  $\mu_S=0.4$. 
It is further assumed that superconductivity is induced into 
the Weyl semimetal by virtue of proximity effects.
From the component of the normalized current normal
 to the interfaces, $J_z$,
the charge current flowing through the junction 
corresponds to $2e J_z|\Delta| W/\hbar$. 
As seen from the components of the  
Hamiltonian in Eq.~(\ref{Hk}), the  conical tilting parameter $\beta$ and the inversion symmetry breaking terms $\alpha_{xy},\alpha_z$
are expected to strongly influence the flow of charge through the junction. 

We show in Fig.~\ref{fig4}, the normalized supercurrent $J_z$ as a function of 
 $\varphi$ for a moderate range of the Weyl cone tilting parameter $\beta$ (see legend).
Two different anisotropic states are considered:
 (a) where  the orbital term vanishes $\alpha_z=0$, and
(b) 
$\alpha_z=0.2$.  
When the orbital term is set to zero, 
we see in Fig.~\ref{fig4}(a) that 
 the supercurrent profile has the conventional sinusoidal $2\pi$ periodicity. 
 Increasing the tilt
of the Weyl cones is shown to cause a non-monotonic variation in the amplitudes. 
Indeed, by tuning $\beta$ appropriately, the current can almost
vanish or change direction altogether. Note that the current-phase 
periodicity changes to $4\pi$ by the presence of higher harmonics 
at the current reversal points \cite{AlidoustBP1,AlidoustBP2}. 
The appearance of higher order harmonics in the 
current-phase-relation can be made more pronounced  by
manipulating the WS parameter set (not shown).
Next, in Fig.~\ref{fig4}(b) 
the inversion symmetry breaking term takes a representative value corresponding to $\alpha_z=0.2$.
By introducing a $z$-component to the inversion symmetry breaking term, the current-phase-relation undergoes a phase shift
leading to a nonvanishing current at $\varphi=0,\pi$, creating a $\varphi_0$-state.
Our results are consistent with recent findings in  quasiclassical Weyl semimetal systems containing  disorder \cite{AlidoustWS2}. 
Within a quasiclassical approach, 
it was found that supercurrent can flow through a triplet channel, and the  $\varphi_0$ state is independent of the diffusion constant, 
demonstrating that this self-biased current can occur regardless of the density of nonmagnetic impurities and disorder in WS systems. 
Hence, the predicted robust self-biased 
current should be experimentally accessible over a wide 
range of parameters.\cite{zu1,zu2,zu3,herve,AlidoustBP1,AlidoustWS2,phi0} 
It is worth mentioning that in a recent experiment, \cite{herve} this self-biased current 
was observed using a $\rm Bi_2 Se_3$ platform, which is in agreement with theoretical predictions 
for topological insulator Josephson junctions \cite{zu1,zu2,zu3}.

We also see that for type-II Weyl semimetals,
 increases in the  conical tilt leads to a nontrivial current that is accompanied by 
the appearance of additional harmonics. 
In Fig.~\ref{fig4}(c), we investigate the effects  of varying the Weyl node separation  parameter $Q$.
Here we set $\alpha_z=0$, and $\beta=0.2$. 
We clearly see that this parameter can also control the direction and magnitude of the supercurrent,
as well as
the appearance of higher harmonics. 
In addition, the junction thickness $d$ can be exploited for controlling these phenomena in 
the supercurrent (not shown). Hence, our results demonstrate that 
a Josephson junction 
made of a type-I or type-II Weyl semimetal offers 
experimentally accessible parameters that can essentially act as 
control knobs
to tune the supercurrent behavior. In many instances, these parameters are 
externally controllable, e.g., by applying an external strain \cite{AlidoustBP1,AlidoustBP2}. For the exhibited  sign reversals, manifestation of higher harmonics, and induction of a
$\varphi_0$ state in the supercurrents shown above,  no 
exchange interaction or Zeeman field was involved. 
This is in contrast to previously proposed 
platforms where the exchange field is an essential ingredient, and
which might complicate experimental situations.    

\begin{figure}[t]
\includegraphics[ width=8.50cm,height=10.5cm]{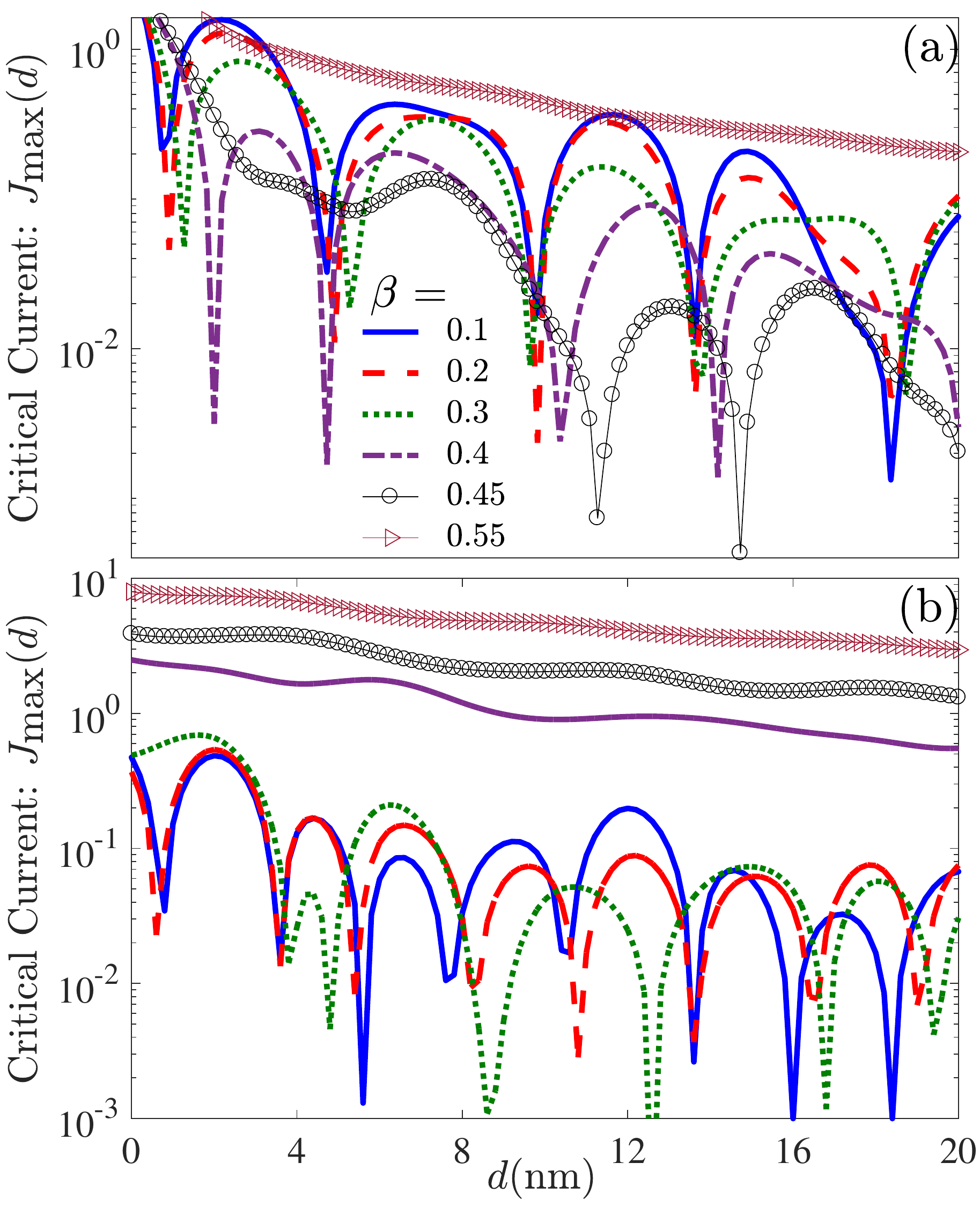}
\caption{ (Color online). Critical normalized charge current $J_\text{max}$ flowing across the Josephson junction shown in Fig. \ref{fig4} as a function of junction thickness $d$. In (a) we set $Q=0.4\pi$ and (b) $Q=0.7\pi$ and show the influence of the tilting parameter $\beta$ on the critical supercurrent. The remaining junction parameters are set as follows: $\gamma=0.5$, $\alpha_{x,y}=1$, $\alpha_{z}=0$, and  $\eta=1$.
 }
\label{fig5}
\end{figure}

An experimentally relevant quantity is the critical supercurrent (maximum of the charge supercurrent defined by $max(|J_z(\varphi)|)$) that can vary when one of the junction parameters changes. The generic behavior of the critical supercurrent in the Weyl semimetal Josephson junction we considered can be inferred from Fig.~\ref{fig4}. For example, from Fig.~\ref{fig4}(a) it is clear that the critical current in such a junction oscillates as a function of $\beta$. 
The behavior of the critical current can similarly be seen by considering other parameters, such as $Q$ (in Fig.~\ref{fig4}(c)). 
These findings are summarized in the critical current plots shown in Fig. \ref{fig5}. To be consistent with the current phase relations presented in Fig. \ref{fig4}, we use identical parameters in Fig. \ref{fig5}. In Fig. \ref{fig5}(a) we set $Q=0.4\pi$ and in Fig. \ref{fig5}(b) $Q=0.7\pi$. The critical supercurrent is shown as a function of junction thickness $d$ for various values of the tilting parameter $\beta=0.1, 0.2, 0.3, 0.4, 0.45, 0.55$. We see that the supercurrent undergoes multiple $0$-$\pi$ reversals by varying the junction thickness. Depending on $Q$ and the value for $\gamma$, increasing the tilting parameter beyond a threshold value causes the critical supercurrent to change its behavior drastically, as it reduces to a simple decaying function. This behavior originates from the now-dominate tilting parameter tilting parameter, which is consistent with Ref. \cite{AlidoustWS2}. Increasing $\beta$ results in amplifying the magnitude of the critical supercurrent. It is worth mentioning that we have purposefully chosen the orientation of the
junction length to be along the $z$ direction. Because the 
Hamiltonian in Eq.~(\ref{Hamil}) is not isotropic, the most interesting features of current transport in the model Weyl semimetal 
we consider appears when the current flows in the $z$ direction. This behavior follows from
 the interplay of broken time-reversal and inversion symmetries in this specific direction. 
 From Eq.~(\ref{Hamil}), it is apparent that the interesting 
 phenomena presented in Fig.~\ref{fig4} are absent when the Josephson junction is directed either along $x$ or $y$, so the SN and NS interfaces are oriented along $z$ ($y$ or $x$ orientations, respectively). In particular, 
 the appearance of the $\varphi_0$ self-biased Josephson state is specific to the 
 charge current component flowing in the $z$ direction in our model Hamiltonian [Eq.~(\ref{Hamil})]. 
 Lastly, it is worth mentioning that recent experiments involving $\rm Cd_3As_2$ semimetal Josephson junctions have 
 found $4\pi$ periodic current phase relations \cite{4pi_1,4pi_2}. This $4\pi$ periodic 
 current phase relation is attributed to the presence of topological superconductivity. 
 Note that a $4\pi$ periodic current phase relation can also appear in a ballistic Josephson junction with surface states of a three-dimensional topological insulator, and persist 
 even in the quasiclassical regime where the chemical potential is  large compared to other energy scales\cite{zu3}.  

\section{conclusions}\label{conclusions}
Starting from equal-pseudospin and unequal-pseudospin phonon mediated spin-singlet electron pair couplings, we have studied the effective
superconducting correlations that can exist in a Weyl semimetal system with the 
possible  tilting of the Weyl cones, rendering the WS into type-II phase. 
We have considered a situation where the
quasiparticle momenta are good quantum numbers, 
and derived the components of the
Green's function that allow us to investigate the 
effective symmetry profiles of various superconducting correlations. 
Our results demonstrate that by properly engineering a Weyl semimetal system 
using intrinsic, and
externally controllable parameters, 
the superconducting correlations can change 
symmetries from effective $s$-wave and $p$-wave classes to $d$-wave and $f$-wave types. 
Our results may provide a theoretical 
scenario for a recent experiment where 
it was observed that superconductivity in the 
$\rm MoTe_2$ Weyl semimetal depends on 
the applied strain of the system \cite{wylsuperc_exp1}.
By diagonalizing the Bogoliubov-de Gennes Hamiltonian 
and deriving the corresponding wavefunctions \cite{AlidoustBP1,AlidoustBP2}
(without any simplifying approximations\cite{AlidoustWS2}), we also
found  the supercurrent flow within a Josephson junction comprised of a ballistic Weyl semimetal. We 
demonstrated  that a Weyl semimetal Josephson junction presents a practical platform to create
a device that has several tunable parameters which control the direction and periodicity 
of supercurrent flow, 
induce multiple $0$-$\pi$ crossovers in the critical charge current, 
 and generates a $\varphi_0$ Josephson state.

\acknowledgments
M.A. is supported by Iran's National Elites Foundation (INEF).  K.H. is supported in part by ONR and a grant of HPC resources from the DOD HPCMP.

\end{document}